\documentclass[showpacs,aps,pra,twocolumn]{revtex4}

\usepackage{color}
\usepackage{bm}
\usepackage{amsmath}
\usepackage{amssymb}
\usepackage{graphicx}
\usepackage{dcolumn}
\usepackage{graphicx}
\usepackage{epsf}
\usepackage{amsmath}
\usepackage{bm}

\def\half{{\textstyle{1\over 2}}}

\def\half{{\textstyle{\frac12}}}

\newcommand{\dd}{\mathrm{d}}
\newcommand{\ii}{\mathrm{i}}

\begin{document}

\title{Self-Energy Correction to the Hyperfine Splitting for Excited States}

\author{B. J. Wundt}
\author{U. D. Jentschura}

\affiliation{Department of Physics,
Missouri University of Science and Technology,
Rolla, Missouri 65409-0640, USA}

\begin{abstract} 
The self-energy corrections to the hyperfine splitting 
is evaluated for higher excited states in
hydrogenlike ions, 
using an expansion in the binding parameter $Z\alpha$,
where $Z$ is the nuclear charge number, and $\alpha$
is the fine-structure constant.
We present analytic results for $D$, $F$ and $G$ 
states, and for a number of highly excited 
Rydberg states with principal quantum numbers 
in the range $13 \leq n \leq 16$,
and orbital angular momenta $\ell = n-2$ and $\ell = n-1$.
A closed-form, analytic expression
is derived for the contribution of high-energy photons,
valid for any state with $\ell \geq 2$ and arbitrary $n$,
$\ell$ and total angular momentum $j$.
The low-energy contributions are written in the 
form of generalized Bethe logarithms and evaluated 
for selected states.
\end{abstract}


\maketitle

% \tableofcontents

%
% Introduction
%
\section{INTRODUCTION}

The self-energy correction to the hyperfine splitting is the 
dominant quantum electrodynamic (QED) correction to the 
magnetic interaction of the bound electron with the field of the nucleus.
The hyperfine interaction energy 
of electron and nucleus is proportional to 
$g_N \, \alpha (Z\alpha)^3 m_e^2/m_N$,
where $g_N$ is the nuclear $g$ factor, and
$m_e$ and $m_N$ are the electron and nuclear masses, respectively.
Relativistic corrections enter at relative order $(Z\alpha)^2$.
The dominant QED correction is due to the 
anomalous magnetic moment of the electron and 
enters at relative order $\alpha$.
Here, we consider the QED correction
of order $\alpha \, (Z\alpha)^2$, which is the sum of 
a high- and a low-energy part.
Relativistic corrections to the anomalous magnetic interaction give 
one of the dominant contributions to the high-energy part,
which can otherwise be calculated on the basis of 
a form-factor approach, using a generalized Dirac equation 
in which the radiative effects and the hyperfine 
interaction are inserted ``by hand.''
The low-energy part constitutes a
correction to the Bethe logarithm due to the 
hyperfine interaction. It can be formulated 
as a hyperfine correction to the self-energy,
the effect being equivalent 
to the self-energy correction to the hyperfine splitting 
mediated by low-energy virtual photons
[up to order $\alpha\,(Z\alpha)^2$].

In our treatment, we follow the formalism of nonrelativistic QED (NRQED)
detailed in Ref.~\cite{CaLe1986}, and refer to
Refs.~\cite{Zw1961,BrPa1967,YeArShPl2005,JeYe2006,YeJe2008,YeJe2010} for a number of
previous investigations regarding the treatment of the self-energy correction
to the hyperfine splitting in systems with low nuclear charge number.

Our paper is organized as follows. The general formalism of the hyperfine
interaction is described in Sec.~\ref{form}.  For the self-energy
correction, the low-energy part is treated in Sec.~\ref{LEP}, and the
high-energy part is calculated in Sec.~\ref{HEP}.  Results and theoretical
predictions are discussed in Sec.~\ref{resu}.  Conclusions are reserved for
Sec.~\ref{conclu}.  Natural units with $\hbar = c = \epsilon_0 = 1$ are used
throughout the paper.

%
% Formalism
%
\section{FORMALISM}
\label{form}

Following the derivation in Ref.~\cite{JeYe2010},
the magnetic dipole field of the
nucleus is described by the vector potential
\begin{equation}\label{Ahfs}
\vec{A}_{\mathrm{hfs}} (\vec{x}) = 
-\frac{1}{4 \pi} \frac{\vec{\mu} \times \vec{x}}{r^3} \,,
\end{equation}
where $\vec x$ is the coordinate vector and $r = | \vec x|$.
The curl of this vector potential yields the magnetic field
\begin{equation}\label{Bhfs}
\vec{B}_{\mathrm{hfs}} = 
\vec{\nabla} \times \vec{A}_{\mathrm{hfs}} (\vec{x}) = 
-\frac23 \; \vec{\mu} \; \delta (\vec{x}) - 
\frac{3 ( \vec{\mu} \cdot \hat{\vec{x}}\, ) 
\hat{\vec{x}} - \vec{\mu} }{4 \pi r^3} \,,
\end{equation}
and the fully relativistic hyperfine interaction Hamiltonian
thus reads
\begin{equation}\label{relhfshamil}
H_{\mathrm{hfs}} = 
-e \vec{\alpha} \cdot \vec{A}_{\mathrm{hfs}} (\vec{x}) = 
\frac{e}{4 \pi} \vec{\alpha} \cdot \frac{\vec{\mu} \times \vec{x}}{r^3} = 
\frac{e}{4 \pi} \vec{\mu} \cdot \frac{\vec{x} \times \vec{\alpha}}{r^3}
\end{equation}
The hyperfine interaction couples 
Dirac eigenstates to the magnetic field of the 
nucleus. The electronic states can be written as 
$|n j m\rangle \equiv |n j \ell m\rangle$,
where $n$ is the principal quantum number,
and the orbital and total angular momenta of the 
electron ($\ell$ and $j$, respectively) can be mapped to the 
Dirac angular quantum number
$\kappa = (-1)^{j-\ell+\half} (j+\half)$.
Finally, $m$ is the 
projection of the total electron angular momentum 
onto the quantization axis.
In this article, we sometimes suppress the 
orbital angular momentum $\ell$ in the notation 
because we consider the coupling of the total 
electron angular momentum $j$ to the nuclear spin.
Nuclear states are denoted as
$|I M \rangle$, where $I$ is the nuclear spin
and $M$ its projection onto the quantization axis.
They are coupled to the electron 
eigenstates $|n j m\rangle$ by the 
hyperfine interaction, to form states with 
quantum number $|n f m_f I j \rangle$
which are eigenstates of the total Dirac$+$hyperfine Hamiltonian
($f$ is the total electron$+$nuclear angular momentum,
and $m_f$ is its projection).
Using Clebsch--Gordan coefficients $C_{I M j m}^{f m_f}$,
the $|n f m_f I j \rangle$ states can be written as 
\begin{equation}
\left| n f m_f I j \right> = 
\sum_{M, m} C_{I M j m}^{f m_f} \left| I M \right> \left| n j m \right>  \,. 
\end{equation}
The hyperfine energy $\Delta E_{\mathrm{hfs}}$ thus reads
\begin{equation}
E_{\mathrm{hfs}} = 
\left< n f m_f I j \left| H_{\mathrm{hfs}} \right| n f m_f I j \right> \,.
\end{equation}
Using the Wigner--Eckhart theorem, the
hyperfine energy  can be rewritten as
\begin{align}
E_{\mathrm{hfs}} =& \;
\alpha \frac{g_N}{2} \frac{m_e}{m_N} 
\left[ f(f+1) - I(I+1) - j(j+1) \right]  
\nonumber\\[2ex]
& \; \times \left< n j \half \left| \frac{\left[\vec{x} \times 
\vec{\alpha} \right]_0}{m_e r^3} \right| n j \half \right>  \,,
\end{align}
where $| n j \half \rangle$ is the Dirac eigenstate with 
a definite angular momentum projection $+\half$,
and $\left[\vec{x} \times \vec{\alpha} \right]_0$ is the 
$z$ component (zero component in the spherical basis) 
of the indicated vector product.

We have thus separated the nuclear from the electronic
variables. A detailed analysis of the separation of the nuclear variables
can also be found in Ref.~\cite{BrPa1967}. This procedure allows to reduce the
evaluations of the hyperfine structure and corrections to it, to the evaluation
of matrix elements of operators acting solely on electronic states.
Specifically, we consider corrections to the 
state-dependent electronic matrix element $\Theta_e$, where
\begin{equation}
\Theta_e  = 
\left< n j \half \left| \frac{\left[\vec{x} \times \vec{\alpha} \right]_0}{m_e r^3} 
\right| n j \half \right> \,.
\end{equation}
The hyperfine interaction energy thus is
\begin{equation}
\label{Ehfs}
E_{\mathrm{hfs}} = 
\alpha \frac{g_N}{2} \frac{m_e}{m_N} 
\left[ f\!(f\!+\!1) - I(I\!+\!1) - j(j\!+\!1) \right] \Theta_e \,.
\end{equation}
Relativistic atomic theory leads to the following result for 
$\Theta_e$ (see Refs.~\cite{PyPaIn1973,JeYe2006})
\begin{equation}
\label{thetarel}
\Theta_e = (Z \alpha)^3 m_e 
\frac{\kappa( 2 \kappa (\gamma +n-|\kappa|) -N)}%
{N^4 \left(\kappa^2 - \tfrac14\right)\gamma (4 \gamma^2-1)} \,,
\end{equation}
where $\gamma=\sqrt{ \kappa^2 - (Z \alpha)^2}$.
The effective principal quantum number 
is $N=\sqrt{(n-|\kappa|)^2 + 2 (n-|\kappa|) \gamma + \kappa^2}$.

%
% Low-Energy Part
%
\section{LOW-ENERGY PART}
\label{LEP}

%
% General Formalism
%
\subsection{General Formalism}

In order to treat low-energy virtual photons,
we apply a Foldy-Wouthuysen transformation to the total
Hamiltonian $H_t$ 
which is the sum of the Dirac--Coulomb Hamiltonian and 
the relativistic hyperfine
interaction Hamiltonian,
\begin{equation}
H_t = H_D + H_{\mathrm{hfs}} = 
\vec{\alpha} \cdot \vec{p} + \beta m_e - 
\frac{Z \alpha}{r} - 
e \, \vec{\alpha} \cdot \vec{A}_{\mathrm{hfs}} (\vec{x}) \,.
\end{equation}
The Foldy-Wouthuysen transformation of this Hamiltonian is carried out as
described in Refs.~\cite{FoWu1950,Zw1961,BjDr1966,JeYe2010,Pa2004}. 
The only difference to the case of the ordinary Dirac 
Hamiltonian is that the odd operator $\mathcal{O}$ used in the 
construction of the transformation~\cite{BjDr1966} now reads 
\begin{equation}
\mathcal{O} = 
\vec{\alpha} \cdot \vec{p} - 
e \, \vec{\alpha} \cdot \vec{A}_{\mathrm{hfs}} (\vec{x}) 
\end{equation}
instead of $\vec{\alpha} \cdot \vec{p}$.
The result of the transformation,
\begin{equation}
H_t' +U H_t U^{-1} = H_{\mathrm{FW}} + H_{\mathrm{HFS}} \,,
\end{equation}
is the sum of the Foldy-Wouthuysen Hamiltonian $H_{\mathrm{FW}}$ 
from Ref.~\cite{FoWu1950}, and
$H_{\mathrm{HFS}}$ is the nonrelativistic hyperfine splitting 
Hamiltonian~\cite{BrPa1967,JeYe2010}
\begin{equation}
\label{nrhfshamil}
H_{\mathrm{HFS}} = \frac{e\, m_e}{4 \pi} \vec{\mu} \cdot \vec{h} = 
\frac{e\, m_e}{4 \pi} \vec{\mu} \cdot 
\left( \vec{h}_S + \vec{h}_D + \vec{h}_L \right) \,.
\end{equation}
It consists of the three parts,
\begin{subequations}\label{nrhfsterms}
\begin{align}
\vec{h}_S &= \frac{4 \pi}{3 m_e^2} \vec{\sigma} \delta (\vec{x}) \,, \\[1ex]
\vec{h}_D &= \frac{3 ( \vec{\sigma} \cdot \hat{x}) \hat{x} - 
\vec{\sigma} }{2 m_e^2 r^3} \,, \\[1ex]
\vec{h}_L &= \frac{\vec{\ell}}{m_e^2 r^3} \,,
\end{align}
\end{subequations}
whose designation is inspired by the apparent angular momentum association of
the terms.  Following the notation in Ref.~\cite{JeYe2010}, lowercase letters
in the subscript are used to label relativistic operators, whereas
nonrelativistic operators are denoted by uppercase letters in the subscript.
However, we use the lowercase notation for the scaled vector quantity
$\vec{h}$ in order to denote the electronic operators
in the nonrelativistic hyperfine Hamiltonian.  
Furthermore, $\hat{x} = \vec x / | \vec x|$ is the
position unit vector.  The zero component ($z$~component) $h_0 = {h}_{S,0} +
{h}_{D,0} + {h}_{L,0}$ of the Hamiltonian ``vector'' $\vec h$ therefore reads
as
\begin{equation}
h_0 = \frac{4 \pi}{3 m_e^2} \sigma_0 \, \delta(\vec{x}) 
+ \frac{3 ( \vec{\sigma} \cdot \hat{x}\, ) \hat{x}_0 - \sigma_0 }{2 m_e^2 r^3} 
+ \frac{\ell_0}{m_e^2 r^3} \,.
\end{equation}
With the help of $h_0$,
the nonrelativistic limit of Eq.~\eqref{thetarel} is obtained as
\begin{equation}\label{thetanr}
\Theta_e^{\mathrm{NR}} = 
\left< n j \ell \half \left| h_0 \right| n j \ell \half \right> = 
\frac{\kappa}{|\kappa|} \,
\frac{ (Z \alpha)^3 m_e}{n^3 ( 2 \kappa+1) (\kappa^2 -\frac14)} \,,
\end{equation}
which we use to define the nonrelativistic quantity 
\begin{align}
E_F &= E_{\mathrm{HFS}}^{\rm NR} = E_{\mathrm{HFS}} =\;
\alpha \frac{g_N}{2} \frac{m_e}{m_N}
\\[2ex]
& \; \times 
\left[ f\!(f\!+\!1) - I(I\!+\!1) - j(j\!+\!1) \right] 
\Theta_e^{\rm NR} \,,
\nonumber
\end{align}
which is commonly referred to as the Fermi energy. 
The relativistic and QED corrections 
can be expressed as multiplicative corrections of
$\Theta_e$, via the replacement
\begin{equation}
\Theta_e^{\mathrm{NR}} \rightarrow
\Theta_e^{\mathrm{NR}} \, \left[ 1 + 
\delta \Theta_e^{\rm rel} + \delta \Theta_e^{\rm QED}
\right] \,.
\end{equation}
By expanding $\Theta_e$ to second order in $Z\alpha$,
we obtain
\begin{align}
\delta \Theta_e^{\mathrm{rel}} =& \;
(Z \alpha)^2 
\left( \frac{12 \kappa^2 -1}{2 \kappa^2 (2 \kappa -1) (2 \kappa +1)} + 
\frac{3}{2 n} \frac{1}{|\kappa|} \right. 
\nonumber\\[2ex]
& \left. + \frac{3-8 \kappa}{2n^2(2 \kappa -1)} \right) 
\end{align}
and the corresponding energy shift
\begin{equation}
\delta E_{\mathrm{hfs}}^{\mathrm{rel}} = 
E_{\mathrm{HFS}} \; \delta \Theta_e^{\mathrm{rel}} \,.
\end{equation}
The QED term
\begin{equation}
\delta E_{\mathrm{HFS}}^{\rm QED} = 
E_{\mathrm{HFS}} \; \delta \Theta_e^{\rm QED} 
\end{equation}
is the subject of this paper. For the QED
corrections terms up to relative order $\alpha (Z \alpha)^2$ with respect to
the nonrelativistic hyperfine splitting will be considered.
In order to do the calculation,
we need the three terms from Eq.~\eqref{nrhfshamil} and 
a further correction to the electron's transition current,
due to the hyperfine interaction.
Namely, in the presence of the hyperfine interaction,
the kinetic momentum of the electron finds a modification 
\begin{equation}
\frac{\vec p}{m_e} \rightarrow \frac{\vec p}{m_e} - 
\frac{e}{m_e} \vec{A}_{\mathrm{hfs}} = 
\frac{\vec p}{m_e} + \frac{|e| \, m_e}{4 \pi} \,
|\vec{\mu}| \; \delta \vec{j}_{\mathrm{HFS}} \,.
\end{equation}
The current
\begin{equation}
\delta \vec{j}_{\mathrm{HFS}} = 
\frac{\hat{\vec{\mu}} \times \vec{x}}{m_e^2 r^3} 
\end{equation}
has the zero component
\begin{equation}
\delta \vec{j}_{0,\mathrm{HFS}} = 
\frac{1}{m_e^2 r^3} \left( -y \, \hat{e}_x + x \, \hat{e}_y \right) \,,
\end{equation}
which is used in the calculations below.

%
% Specific Terms
%
\subsection{Specific Terms}

Following Ref.~\cite{JeYe2010}, there are four corrections, which arise from the
correction of the interaction current, from the correction of the Hamiltonian,
from the correction of the reference state energy, and finally from the
correction of the reference-state wave function. 
We first treat the 
hyperfine correction to the interaction current
and to this end, define a useful normalization factor
\begin{equation}
\mathcal{N} = 
\frac{1}{\left< n j \ell \half \left| h_0 \right| n j \ell \half \right>} = 
\frac{1}{\Theta_e^{\mathrm{NR}}} \,.
\end{equation}
The hyperfine correction to the interaction current
is then given as
\begin{align}
& \delta \Theta_L^{\delta j}  = 
\frac{4 \alpha \mathcal{N}}{3 \pi} \int\limits_0^{\epsilon} \dd\omega_{\vec{k}} 
\omega_{\vec{k}} \!\! \sum_{n'j'\ell' m'} \!\!
\left< n j \ell \half \left| \frac{p^i}{m_e} \right| n' j' \ell' m' \right>  
\nonumber\\
& \; \times \frac{1}{E_n -E_{n'} - 
\omega_{\vec{k}}} 
\left< n' j' \ell' m' \left| \delta j^i_{0,\mathrm{HFS}} 
\right| n j \ell \half \right> 
\nonumber\\[2ex]
& = \frac{\alpha}{\pi} (Z \alpha)^2 \frac{4 \mathcal{N}}{3 (Z \alpha)^2} 
\sum_{n'j'\ell' m'} (E_n -E_{n'}) \ln \left( \frac{|E_{n'} - E_n|}{m_e (Z \alpha)^2} 
\right) 
\nonumber\\
& \times \left< n j \ell \half \left| \frac{p^i}{m_e} \right| n' j' \ell' m' \right> 
\left< n' j' \ell' m' \left| \delta j^i_{0,\mathrm{HFS}} \right| n j \ell \half 
\right> .
\end{align}
The term containing the logarithm of $\epsilon$,
which is a scale-separation parameter that cancels when 
high- and low-energy parts are added~\cite{Pa1993}, 
vanishes after angular integration in the matrix element. The structure of the
logarithmic term here is very similar to the Bethe logarithm encountered in
Ref.~\cite{Be1947}. Terms of this form will arise for the other corrections in the
low-energy part as well. In the following, these terms are denoted as
$\beta_{\mathrm{HFS}}$ and are evaluated numerically with the
methods described in Ref.~\cite{SaOe1989}. Thus, the low-energy correction due
to the nuclear-spin dependent current is
\begin{equation}
\delta \Theta_L^{\delta j} = 
\frac{\alpha}{\pi} (Z \alpha)^2 
\beta_\mathrm{HFS}^{\delta j} \,.
\end{equation}

Next, we treat the
corrections to the Hamiltonian, to the energy and to the wave function.
The perturbation due to the 
hyperfine splitting Hamiltonian yields the term
[we define the resolvent 
$G(\omega_{\vec{k}}) = 1/(E_n -H_S - \omega_{\vec{k}})$]
\begin{align}
& \delta \Theta_L^{\delta H}  = 
\frac{2 \alpha \mathcal{N}}{3 \pi} \int_0^{\epsilon} \dd \omega_{\vec{k}} \; 
\omega_{\vec{k}} 
\nonumber\\ 
& \times 
\left< n j \ell \half \left| \frac{p^i}{m_e} 
G(\omega_{\vec{k}}) \, h_0 \, G(\omega_{\vec{k}}) \frac{p^i}{m_e} 
\right| n j \ell \half \right> 
\nonumber\\[2ex]
& = \frac{2 \alpha \mathcal{N}}{3 \pi m_e^2} 
\ln \left[ \frac{\epsilon}{m_e (Z \alpha)^2} \right] 
\nonumber\\[2ex]
& \times \left< n j \ell \half 
\left| \left(  \frac12 [p^i,[h_0,p^i]] + p^2 h_0 \right) \right| n j \ell \half \right> 
\nonumber\\[2ex]
& \quad \quad +\frac{\alpha}{\pi} 
(Z \alpha)^2 \beta_\mathrm{HFS}^{\delta H} \,,
\end{align}
The correction to the energy denominator 
in the Schr\"{o}dinger propagator can be written as
\begin{align}
& \delta \Theta_L^{\delta E} = 
-\frac{2 \alpha \mathcal{N}}{3 \pi} 
\int_0^{\epsilon} d \omega_{\vec{k}} \; \omega_{\vec{k}} 
\left< n j \ell \half \left| h_0 \right| n j \ell \half \right> 
\nonumber\\
& \qquad \times 
\left< n j \ell \half \left| 
\frac{p^i}{m_e} \left[ G(\omega_{\vec{k}}) \right]^2  
\frac{p^i}{m_e} \right| n j \ell \half \right> 
\nonumber\\
&= - \frac{2 \alpha \mathcal{N}}{3 \pi m_e^2} 
\ln \left[ \frac{\epsilon}{m_e (Z \alpha)^2} \right] 
\left< n j \ell \half \left| p^2 \right| n j \ell \half \right> 
\nonumber\\
& \qquad \times \left< n j \ell \half \left| h_0 \right| n j \ell \half \right> 
+ \frac{\alpha}{\pi} (Z \alpha)^2 \beta_\mathrm{HFS}^{\delta E} \,,
\end{align}
where the prime indicates the reduced Green function.
Finally, the correction to the wave function due to the hyperfine splitting
Hamiltonian is
\begin{align}
& \delta \Theta_{L}^{\delta \Phi} = 
\frac{4 \alpha \mathcal{N}}{3 \pi} 
\int_0^{\epsilon} d \omega_{\vec{k}} \; \omega_{\vec{k}} 
\nonumber\\ 
& \times 
\left< n j \ell \half \left| 
\frac{p^i}{m_e} G(\omega_{\vec{k}}) \frac{p^i}{m_e} 
\left( \frac{1}{E_n -H_S} \right)' h_0 \right| n j \ell \half \right> 
\nonumber\\ 
&= \frac{4 \alpha \mathcal{N}}{3 \pi m_e^2} \,
\ln \!\! \left[ \frac{\epsilon}{m_e (Z \alpha)^2} \right] 
\nonumber\\ 
& \quad \times \left< n j \ell \half \left| p^i (H_S \!-\! E_n) p^i 
\left( \frac{1}{E_n\! -\!H_S} \! \right)' \! h_0 \right| n j \ell \half \right> 
\nonumber\\ 
& \quad \quad + 
\frac{\alpha}{\pi} (Z \alpha)^2 \beta_\mathrm{HFS}^{\delta \Phi} \,.
\end{align}
Using commutator relations, one can finally
sum up all four corrections in the low-energy part to
\begin{align}
&\delta \Theta_{L} = \delta \Theta_L^{\delta j} 
+ \delta \Theta_L^{\delta H} 
+ \delta \Theta_L^{\delta E} 
+ \delta \Theta_L^{\delta \Phi} 
\nonumber\\
& = 
\frac{\alpha \mathcal{N}}{3 \pi m_e^2} 
\ln \left[ \frac{\epsilon}{m_e (Z \alpha)^2} \right] 
\left< n j \ell \half \left| [p^i,[h_0,p^i]] \right| n j \ell \half \right> 
\nonumber\\
& \qquad + \frac{\alpha}{\pi} (Z \alpha)^2 \beta_{\mathrm{HFS}} \,,
\end{align}
where $\beta_{\mathrm{HFS}} $ is the sum
\begin{equation}
\label{betahfs}
\beta_{\mathrm{HFS}} = 
\beta_\mathrm{HFS}^{\delta j} + 
\beta_\mathrm{HFS}^{\delta H} + 
\beta_\mathrm{HFS}^{\delta E} + 
\beta_\mathrm{HFS}^{\delta \Phi} \,.
\end{equation}
The double commutator 
\begin{equation}
\left< n j \ell \half \left| [p^i,[h_0,p^i]] \right| n j \ell \half \right> 
= \left< n j \ell \half \left| \vec\nabla^2 h_0 \right| n j \ell \half \right>
\end{equation}
vanishes for states with $\ell \geq 2$ up to and 
including order $(Z \alpha)^5$,
and hence $\delta \Theta_{L}$ takes the very simple
form
\begin{equation}
\label{ThetaL}
\delta \Theta_{L} = 
\frac{\alpha}{\pi} (Z \alpha)^2 \beta_{\mathrm{HFS}} \,.
\end{equation}

%
% High-Energy Part
%
\section{HIGH-ENERGY PART}
\label{HEP}

Up to relative order  $\alpha (Z \alpha)^2 E_F$,
it is sufficient~\cite{JeYe2010} to consider the problem on the level of the 
modified Dirac Hamiltonian
\begin{align}
H_D^{(m)} =& \; \vec{\alpha} \left[ \vec{p} 
-e F_1(\vec \nabla^2) \vec{A} \right] 
+ \beta m_e + F_1(\vec \nabla^2) V 
\nonumber\\[2ex]
& \; + F_2(\vec\nabla^2) \frac{e}{2m_e} 
\left( \ii \vec{\gamma} \cdot \vec{E} - \beta \,\vec{\Sigma} \cdot \vec{B} \right) \,,
\end{align}
where $F_1$ and $F_2$ are the one-loop Dirac and Pauli form factors
of the electron, respectively. Their expressions are known
(see Chapter 7 of Ref.~\cite{ItZu1980}).

The $F_1$ form factor slope gives rise to the following
effective interaction
\begin{equation}
-e F_1'(0) \vec \nabla^2 \vec{\alpha} \cdot \vec{A}_{\mathrm{hfs}} = 
\frac{\alpha}{3 \pi} \left[ \ln \left( \frac{m_e}{2 \epsilon} \right) + 
\frac{11}{24} \right] \vec \nabla^2 H_{\mathrm{hfs}} \,.
\end{equation}
Up to the order $\alpha(Z\alpha)^2 E_F$, 
we may write the correction in terms of the nonrelativistic 
hyperfine Hamiltonian $h_0$,
\begin{align}
\delta \Theta_{H,1} = 
\frac{\alpha \mathcal{N}}{3 \pi m_e^2} 
\left[ \ln \left( \frac{m_e}{2 \epsilon} \right) + \frac{11}{24} \right] \,
\left< n j \ell \half \left| \vec\nabla^2 h_0 \right| n j \ell \half \right> \,.
\end{align}
However, as already pointed out,
the matrix element of $\vec\nabla^2 h_0$ 
vanishes on states 
with $\ell \geq 2$ which are relevant to our investigations,
and so
\begin{equation}
\delta \Theta_{H,1} = 0 \,.
\end{equation}
The second correction is a second-order perturbation
involving the $F_1$ correction to the Coulomb potential,
\begin{align}
\delta \Theta_{H,2} = & \;
\frac{2 \alpha \mathcal{N}}{3 \pi m_e^2} 
\left[ \ln \left( \frac{m_e}{2 \epsilon} \right) + \frac{11}{24} \right] 
\\[2ex]
& \; \times 
\left< n j \ell \half \left| \vec\nabla^2 V \left( \frac{1}{E_n -H_S} \right)' h_0 
\right| n j \ell \half \right> \,.
\nonumber
\end{align}
Again, $\vec\nabla^2 V$ is proportional to the Dirac $\delta$ and 
therefore vanishes for states with $\ell\geq 1$. 
Accordingly, for states with $\ell \geq 2$ we have
\begin{equation}
\delta \Theta_{H,2} = 0 \,.
\end{equation}

The Pauli $F_2$ form factor gives rise to a second-order perturbation 
involving a magnetic moment correction to the Coulomb potential,
\begin{align}
\delta \Theta_{H,3} = & \;
2 \mathcal{N} \, F_2 (0) 
\\
& \; \times \left< \psi^{\dagger} \left| \frac{-\ii}{2 m_e} 
\vec{\gamma} \cdot \vec{\nabla} V \left( \frac{1}{E_{\psi} -H_D} \right)' 
H_{\mathrm{hfs}} \right| \psi \right> \,,
\nonumber
\end{align}
where $F_2$ is the magnetic form factor. For $F_2 (0)$, the Schwinger value
$F_2 (0)= \frac{\alpha}{2 \pi}$ may be used. 
After a Foldy--Wouthuysen transformation, we can write
$\delta \Theta_{H,3}$ as the sum of two terms.
The first of these, $\delta \Theta_{H,3n}$, involves no
mixing of upper and lower components in the Dirac wave function
and reads
\begin{align}
& \delta \Theta_{H,3n} = 
\frac{\alpha \mathcal{N}}{2 \pi m_e^2} 
\nonumber\\[2ex]
& \; \times 
\left< n j \ell \half \left| \frac{Z \alpha}{r^3} \vec{\sigma} \cdot \vec{\ell} 
\left( \frac{1}{E_{n} -H_S} \right)' H_{\mathrm{HFS}} \right| n j \ell \half \right> \,.
\end{align}
We find the following general result for states
with $\ell \geq 2$,
\begin{align}
& \delta \Theta_{H,3n} = 
\frac{\alpha}{\pi} (Z \alpha)^2 \left( \frac{1}{2 \ell \kappa} \; 
\frac{60 \ell^4+120 \ell^3+55 \ell^2-5 \ell-3}{(2 \ell+1)^2 \, 
(4\ell^3 +8 \ell^2 + \ell-3)} \right. 
\nonumber\\
& \qquad \left. 
+ \frac{3}{2 n} \frac{1}{\kappa (2\ell +1)} - 
\frac{3}{8 n^2} \frac{\ell ( \ell+1)}{\kappa ( \ell+\frac32)( \ell- \frac12)} 
\right) \,.
\end{align}
Lower components of the Dirac wave function give rise to the 
mixing term
\begin{align}
\delta \Theta_{H,3m} = & \;
- \ii \frac{\alpha \mathcal{N}}{2 \pi m_e} 
\left< n j \ell \half \left| \frac{Z \alpha}{r^3} 
\left( \vec{\gamma} \cdot \vec{x} \right) 
\frac{1}{2m_e} H_{\mathrm{hfs}} \right| n j \ell \half \right> 
\nonumber\\[2ex]
=& \; \frac{\alpha \mathcal{N}}{ \pi}  
\frac{- \kappa}{8 j (j+1)} \,
\left< n j \ell \half \left| \frac{Z \alpha}{m_e^3 r^4} 
\right| n j \ell \half \right> \,.
\end{align}
We find the general result
\begin{align}
\delta \Theta_{H,3m} = & \;
\frac{\alpha}{\pi} (Z \alpha)^2 \frac{|\kappa|}{4 j (j+1)} \frac{(2 \kappa +1) 
(\kappa^2 - \frac14)}{(2 \ell-1) (2 \ell+3) ( \ell + \frac12)} 
\nonumber\\
& \; \times \left( \frac{1}{n^2} - \frac{3}{\ell (\ell+1)} \right) \,.
\end{align}
The $F_2$ correction to the 
magnetic photon exchange of electron and nucleus gives 
rise to the effective interaction
\begin{equation}
\label{F2}
- F_2 (\vec \nabla^2) \frac{e}{2m_e} \beta\, \vec{\Sigma} \cdot \vec{B}_{\mathrm{hfs}} = 
F_2 (\vec \nabla^2) 
\left[ \frac{e m_e}{4 \pi} \beta \vec{\mu} \cdot
\left( \vec{h}_s + \vec{h}_d \right) \right] \,.
\end{equation}
Here, $\vec{h}_s$ and $\vec{h}_d$ are  the generalizations of 
$\vec{h}_S$ and $\vec{h}_D$ to $4\times 4$ matrices,
\begin{subequations}
\begin{align}
\vec{h}_s &= 
\frac{4 \pi}{3 m_e^2} \, \vec{\Sigma} \; \delta (\vec{x}) \,, \\ 
\vec{h}_d &= 
\frac{3 \, ( \vec{\Sigma} \cdot \hat{\vec{x}} ) \, 
\hat{\vec{x}} - \vec{\Sigma}}{2 \,m_e^2\, r^3} \,.
\end{align}
\end{subequations}
Taking $F_2(\vec \nabla^2) \approx F_2(0)$ in Eq.~\eqref{F2}, we obtain 
the correction
\begin{align}
\delta \Theta_{H,4} =& \;
\mathcal{N} F_2(0) \left< \psi \left| \beta \left( h_{s,0} + h_{d,0} \right) 
\right| \psi \right> 
\nonumber\\[2ex]
=& \; 
\frac{\alpha \mathcal{N}}{2 \pi} \,
\left< \psi \left| \beta \left( h_{s,0} + h_{d,0} \right)  \right| \psi \right> \,.
\end{align}
Generalizing results from Ref.~\cite{WuJe2010} for the 
term of relative order $\alpha$, we find the result
\begin{align}
\delta \Theta_{H,4} &= 
\frac{\alpha}{\pi} \left[ \frac{1}{4 \kappa} + 
(Z \alpha)^2 \left( 
\frac{24 \kappa ^3+18 \kappa ^2-\kappa -1}%
{8 \kappa^3(4 \kappa^3+4 \kappa ^2-\kappa -1)} \right. \right. 
\nonumber\\
& \qquad \quad \left. \left. 
+ \frac{3}{8n} \frac{1}{|\kappa| \kappa} 
+ \frac{1}{n^2} \frac{1}{2 \kappa} \frac{1-3\kappa}{2\kappa -1} 
\right) \right] \,.
\end{align}
Taking the slope of $F_2$ in Eq.~\eqref{F2}, we obtain
\begin{align}
& F_2'(0) \frac{e}{2m_e} \beta \, \vec \nabla^2 \, 
\vec{\Sigma} \cdot \vec{B}_{\mathrm{hfs}} 
\nonumber\\
& \qquad = \frac{\alpha}{12 \pi} 
\left[ \frac{e m_e}{4 \pi} \beta \vec{\mu} \cdot 
\left\{ \vec \nabla^2 \left( \vec{h}_s + \vec{h}_d \right) \right\} \right] \,,
\end{align}
with $F_2'(0)=\alpha/12 \pi$. As $F_2'(0) \vec\nabla^2$ already is of 
relative order $\alpha (Z \alpha)^2$, 
this operator only has to be applied to the nonrelativistic wave
function where it vanishes for states 
with $\ell \geq 2$ and thus
\begin{equation}
\delta \Theta_{H,5} = 
\frac{\alpha \mathcal{N}}{12 \pi} 
\left< n j \ell \half \left| 
\vec\nabla^2 \left( h_{S,0} + h_{D,0} \right) 
\right| n j \ell \half \right> = 0\,. 
\end{equation}
%

%
% TABLE BLOCK START
%
\begin{table}[t!]
\caption{\label{table1}
Low-energy contribution $\beta_{\mathrm{HFS}}$ of the self-energy
correction for the hyperfine splitting for $D$ states ($\ell=2$).}
\begin{center}
\begin{tabular}{c@{\hspace{0.2in}}c@{\hspace{0.2in}}c}
\hline
\hline
\rule[-2mm]{0mm}{6mm}
$n$ &
\multicolumn{1}{c}{$\beta_\mathrm{HFS} (n D_{3/2})$} &
\multicolumn{1}{c}{$\beta_\mathrm{HFS} (n D_{5/2})$} \\ 
\hline
\rule[-2mm]{0mm}{6mm}
3 & $- 2.068~39(5) \times 10^{-2}$ & $- 3.455~22(5) \times 10^{-2}$ \\
\rule[-2mm]{0mm}{6mm}
4 & $- 1.302~99(5) \times 10^{-2}$ & $- 3.793~94(5) \times 10^{-2}$ \\
\rule[-2mm]{0mm}{6mm}
5 & $- 1.025~92(5) \times 10^{-2}$ & $- 3.965~95(5) \times 10^{-2}$ \\
\rule[-2mm]{0mm}{6mm}
6 & $- 0.943~28(5) \times 10^{-2}$ & $- 4.084~82(5) \times 10^{-2}$  \\
\hline
\hline
\end{tabular}
\end{center}
\end{table}
\begin{table}[t!]
\caption{\label{table2} Low-energy contribution $\beta_{\mathrm{HFS}}$ of the
self-energy correction for the hyperfine splitting for $F$ states
($\ell=3$).}
\begin{center}
\begin{tabular}{c@{\hspace{0.2in}}c@{\hspace{0.2in}}c}
\hline
\hline
\rule[-2mm]{0mm}{6mm}
$n$ &
\multicolumn{1}{c}{$\beta_{\mathrm{HFS}} (n F_{5/2})$} &
\multicolumn{1}{c}{$\beta_{\mathrm{HFS}} (n F_{7/2})$} \\ 
\hline
\rule[-2mm]{0mm}{6mm}
4 & $- 1.021~46(5) \times 10^{-2}$ & $- 0.953~04(5) \times 10^{-2}$ \\
\rule[-2mm]{0mm}{6mm}
5 & $- 0.683~94(5) \times 10^{-2}$ & $- 1.077~62(5) \times 10^{-2}$ \\
\rule[-2mm]{0mm}{6mm}
6 & $- 0.504~64(5) \times 10^{-2}$ & $- 1.141~28(5) \times 10^{-2}$  \\
\rule[-2mm]{0mm}{6mm}
7 & $- 0.407~32(5) \times 10^{-2}$ & $- 1.182~43(5) \times 10^{-2}$  \\
\hline
\hline
\end{tabular}
\end{center}
\end{table}

\begin{table}[t!]
\caption{\label{table3}
Low-energy contribution $\beta_{\mathrm{HFS}}$ of the self-energy
correction for the hyperfine splitting for $G$ states ($\ell=4$).}
\begin{center}
\begin{tabular}{c@{\hspace{0.2in}}c@{\hspace{0.2in}}c}
\hline
\hline
\rule[-2mm]{0mm}{6mm}
$n$ &
\multicolumn{1}{c}{$\beta_{\mathrm{HFS}} (n G_{7/2})$} &
\multicolumn{1}{c}{$\beta_{\mathrm{HFS}} (n G_{9/2})$} \\ 
\hline
\rule[-2mm]{0mm}{6mm}
5 & $- 0.368~23(5) \times 10^{-2}$ & $- 0.341~76(5) \times 10^{-2}$ \\
\rule[-2mm]{0mm}{6mm}
6 & $- 0.147~39(5) \times 10^{-2}$ & $- 0.388~58(5) \times 10^{-2}$  \\
\rule[-2mm]{0mm}{6mm}
7 & $- 0.012~93(5) \times 10^{-2}$ & $- 0.414~24(5) \times 10^{-2}$  \\
\rule[-2mm]{0mm}{6mm}
8 & $\phantom{-} 0.072~16(5) \times 10^{-2}$ & $- 0.430~98(5) \times 10^{-2}$  \\
\hline
\hline
\end{tabular}
\end{center}
\end{table}

\begin{table*}[th!] 
\begin{center}
\begin{minipage}{14cm}
\begin{center}
\caption{\label{table4} Low-energy contribution $\beta_{\mathrm{HFS}}$ of the 
self-energy correction for the hyperfine splitting for highly excited
states. The numbers in parentheses are standard uncertainties in the last
figure. The exponent of the numerical data is chosen 
to be the same as in Tables~\ref{table1},~\ref{table2}, and~\ref{table3},
illustrating the monotonic decrease of the coeffcients with the 
angular momentum quantum numbers.}
\begin{center}
\begin{tabular}{ c@{\hspace{0.2in}}c@{\hspace{0.2in}}c@{\hspace{0.2in}}c@{\hspace{0.2in}}c@{\hspace{0.3in}}c@{\hspace{0.2in}}c@{\hspace{0.2in}}c@{\hspace{0.2in}}} 
\hline
\hline
\rule[-2mm]{0mm}{6mm} 
$n$ & $\ell$ & $2j$ & $\kappa$ &
\multicolumn{1}{c}{$\beta_{\mathrm{HFS}}$} & $2j$ & $\kappa$ &
\multicolumn{1}{c}{$\beta_{\mathrm{HFS}}$} \\ 
\hline
\rule[-2mm]{0mm}{6mm} 
16  & 15  & $29$  & 15 &  $\phantom{-} 0.006\,310(5)\times 10^{-2}$ & $31$ & -16 &
$- 0.002\,130(5)\times 10^{-2}$ \\ 
\rule[-2mm]{0mm}{6mm} 
16  & 14  & $27$  & 14 &
$\phantom{-} 0.016\,397(5)\times 10^{-2}$ & $29$ & -15 & $- 0.003\,041(5)\times 10^{-2}$ \\
\rule[-2mm]{0mm}{6mm} 
15  & 14  & $27$  & 14 & $\phantom{-} 0.006\,888(5)\times 10^{-2}$ & $29$ & -15 &
$- 0.002\,795(5)\times 10^{-2}$  \\ 
\rule[-2mm]{0mm}{6mm} 
15  & 13  & $25$  & 13 &
$\phantom{-} 0.019\,372(5)\times 10^{-2}$ & $27$ & -14 & $- 0.004\,086(5)\times 10^{-2}$ \\
\rule[-2mm]{0mm}{6mm} 
14  & 13  & $25$  & 13 & $\phantom{-} 0.007\,420(5)\times 10^{-2}$ & $27$ & -14 &
$- 0.003\,741(5)\times 10^{-2}$  \\
\rule[-2mm]{0mm}{6mm} 
14  & 12  & $23$  & 12 &
$\phantom{-} 0.023\,029(5)\times 10^{-2}$ & $25$ & -13 & $- 0.005\,617(5)\times 10^{-2}$ \\ 
\rule[-2mm]{0mm}{6mm}
13  & 12  & $23$  & 12 & $\phantom{-} 0.007\,794(5)\times 10^{-2}$ & $25$ & -13
& $- 0.005\,122(5) \times 10^{-2}$  \\ 
\hline
\hline
\end{tabular} 
\end{center}
\end{center}
\end{minipage}
\end{center}
\end{table*}
%
% TABLE BLOCK END
%

Finally, since
\begin{equation}
\delta \Theta_{H,1} =
\delta \Theta_{H,2} =
\delta \Theta_{H,5} = 0 \,,
\end{equation}
we have for the high-energy part the result
\begin{equation}
\delta \Theta_{H} = \delta \Theta_{H,3n} + \delta \Theta_{H,3m} 
+ \delta \Theta_{H,4} \,.
\end{equation}
It is quite surprising that the 
result obtained by adding the above expressions,
\begin{align}
& \delta \Theta_{H} = 
\frac{\alpha}{\pi} \left\{ \frac{1}{4 \kappa} + 
(Z \alpha)^2 \left[ \frac{1}{8 \kappa^3} \; 
\frac{24 \kappa ^3+18 \kappa ^2-\kappa -1}%
{4 \kappa^3+4 \kappa ^2-\kappa -1} \right. \right. 
\nonumber\\
& + \frac{1}{2 \ell \kappa} \; 
\frac{60 \ell^4+120 \ell^3+55 \ell^2-5 \ell-3}{(2 \ell+1)^2 \, 
(4\ell^3 +8 \ell^2 + \ell-3)} 
\nonumber\\
& - \frac{3}{\ell (\ell+1)} 
\frac{j+\frac12}{4 j (j+1)} \frac{(2 \kappa +1) 
(\kappa^2 - \frac14)}{(2 \ell-1) (2 \ell+3) ( \ell + \frac12)} 
\nonumber\\
& + \frac{1}{n} \, \frac{3}{8 \kappa} \, 
\frac{4 (j+\frac12)+(2\ell+1)}{(2\ell +1) (j+\frac12)} +
\frac{1}{8n^2}  \left( \frac{4}{1-2 \kappa }-\frac{4}{\kappa } \right. 
\nonumber\\
& \left.\left. \left. + 
\frac{(2 j+1) (2 \kappa -1) (2 \kappa +1)^2}%
{2 j (j+1) (2 \ell+3) \left(4 \ell^2-1\right)} -
\frac{3 \ell ( \ell+1)}{\kappa ( \ell+\frac32)( \ell- \frac12)}\right) \right] 
\right\} \,.
\end{align}
can actually be simplified quite considerably,
\begin{align}
\label{ThetaH}
& \delta \Theta_{H} =
\frac{\alpha}{\pi} \left\{ \frac{1}{4 \kappa} + 
(Z \alpha)^2 \left[
\frac{1}{n} \, 
\frac{3}{8} \frac{\kappa}{|\kappa|} \, \frac{6 \kappa+1}{\kappa^2 (2 \kappa+1)} 
\right. \right. 
\\
& \; 
\left. \left. 
+ \frac{1}{n^2} \frac{4 \kappa -1}{2 \kappa (1-2 \kappa)} 
+ \frac{(4 \kappa+1) (6 \kappa +1) (6 \kappa^2+ 3 \kappa -1)}%
{8 \kappa^3 (2 \kappa +1)^2 (2 \kappa -1) (\kappa +1) } 
\right] \right\} \,.
\nonumber
\end{align}
%

%
% Results
%
\section{RESULTS AND PREDICTIONS}
\label{resu}

The total self-energy correction to the hyperfine splitting is 
obtained as the sum of the high- and low-energy parts 
given in Eqs.~\eqref{ThetaH} and~\eqref{ThetaL}, which reads
\begin{align}
\label{hfstotalcorr}
& \delta \Theta = 
\frac{\alpha}{\pi} \left\{ \frac{1}{4 \kappa} \right.
\\
& \;
+ (Z \alpha)^2 \left[ \frac{1}{8 \kappa^3} \; 
\frac{(4 \kappa+1) (6 \kappa +1) (6 \kappa^2+ 3 \kappa -1)}%
{(2 \kappa +1)^2 (2 \kappa -1) (\kappa +1)} \right. 
\nonumber\\
& \left. \left. 
+ \frac{1}{n} \, \frac{3}{8} \frac{\kappa}{|\kappa|} \, 
\frac{6 \kappa+1}{\kappa^2 (2 \kappa+1)} +\frac{1}{n^2} 
\frac{4 \kappa -1}{2 \kappa (1-2 \kappa)} +
\beta_{\mathrm{HFS}}  \right] \right\} \,.
\nonumber
\end{align}
The Bethe logaritm type correction $\beta_{\mathrm{HFS}}$ is
given in Eq.~\eqref{betahfs}.

Restoring the reduced-mass dependence
[we define $r(\mathcal{N}) \equiv m_e/m_N$] 
and adding the relativistic correction of relative order
$(Z\alpha)^2$, we find that 
\begin{align}
\label{vhfscomp}
& \nu_{\mathrm{hfs}} = R_{\infty} c
\frac{Z^3 \alpha^2}{n^3} 
\frac{r(\mathcal{N})}{[1+r(\mathcal{N})]^3} \frac{\kappa}{|\kappa|} 
\frac{g_N}{(2 \kappa +1) (\kappa^2 -\textstyle{\frac{1}{4}})} 
\nonumber\\
& \times 
\left[ f(f+1) - I(I+1) - j(j+1) \right] \times
\biggl\{ 1 + (Z \alpha)^2 
\nonumber\\
& 
\times \left[ 
\frac{12 \kappa^2 -1}{2 \kappa^2 (2 \kappa -1) (2 \kappa +1)} 
+ \frac{3}{2 n} \frac{1}{|\kappa|} 
+ \frac{3-8 \kappa}{2n^2(2 \kappa -1)} \right] 
\nonumber\\
& + \frac{\alpha}{\pi} \frac{1}{4 \kappa} + \frac{\alpha}{\pi} 
(Z \alpha)^2 \left[ 
\frac{3}{8n} \frac{\kappa}{|\kappa|} \, 
\frac{6 \kappa+1}{\kappa^2 (2 \kappa+1)} +\frac{1}{n^2} 
\frac{4 \kappa -1}{2 \kappa (1-2 \kappa)} 
\right.  
\nonumber\\
& \left. 
+ \frac{1}{8 \kappa^3} \; \frac{(4 \kappa+1) 
(6 \kappa +1) (6 \kappa^2+ 3 \kappa -1)}%
{(2 \kappa +1)^2 (2 \kappa -1) (\kappa +1)} 
+\beta_{\mathrm{HFS}} \right] \biggr\} \,.
\end{align}
Numerical data for 
$\beta_{\mathrm{HFS}}$ for $D$, $F$, and $G$ states,
and selected Rydberg states,
can be found in Tables~\ref{table1},~\ref{table2},~\ref{table3},
and~Table~\ref{table4}, respectively.
The latter are relevant for a possible determination
of fundamental constants from Rydberg state
spectroscopy in hydrogenlike ions of medium
nuclear charge number (see Ref.~\cite{JeMoTaWu2008}).

A numerical example:
For the transition $\left| 1 \right> \leftrightarrow \left| 2 \right>$
in atomic hydrogen ($Z=1$) where
\begin{align}
\left| 1 \right> &= \left| n=15, \ell=14, j= 29/2, f=15 \right> \,,\\
\left| 2 \right> &= \left| n=16, \ell=15, j= 31/2, f=16 \right> \,, 
\end{align}
we find the following frequency shift from the Dirac value indicated in
Eq.~(2) of Ref.~\cite{WuJe2010},
\begin{equation}
\Delta \nu_{\mathrm{hfs},1\rightarrow 2} = 
96.759\,8630(8) \, \mathrm{Hz} + 78.764\,693(13) \, \mathrm{Hz} \,,\\
\end{equation}
where the first term is due to the 
hyperfine effects calculated here, and the 
second term is due to relativistic recoil 
and QED effects calculated in Ref.~\cite{JeMoTaWu2008}.
The final theoretical prediction 
for the shift from the Dirac value is 
\begin{equation}
\Delta \nu_{\mathrm{hfs},1\rightarrow 2} = 175.524\,556(13) \, \mathrm{Hz} \,,
\end{equation}
where the fundamental constants of 
CODATA 2006 \cite{MoTaNe2008} 
have been used in the numerical evaluation.
The next higher-order term neglected here is the 
recoil correction of relative order $(Z\alpha)^2\, r(\mathcal N)$,
for which a general expression has been derived in Ref.~\cite{St1963}
[the corresponding expression also is given in Eq.~(42) of Ref.~\cite{JeYe2006}].
The recoil correction is numerically suppressed for $Z=1$.

%
% CONCLUSIONS
%
\section{CONCLUSIONS}
\label{conclu}

Rydberg states of hydrogenlike 
ions with medium nuclear charge number have been proposed as a
device for the determination of fundamental constants~\cite{JeMoTaWu2008}.
Here, we demonstrate that it is possible to obtain accurate 
theoretical predictions for transition frequencies
even in cases where the nucleus carries spin.
To this end, we calculate the self-energy correction to the 
hyperfine splitting of the high-lying states.
Vacuum polarization effects can be neglected for states
with $\ell \geq 2$ to the order relevant for the current
investigation.

We split the calculation into a low-energy part, which 
contains Bethe logarithm type corrections
(Sec.~\ref{LEP}), and a high-energy part, which 
can be treated on the basis of electron form factors
(Sec.~\ref{HEP}).
For the low-energy part, we find that the net result 
can be expressed as the sum of corrections due to the 
hyperfine Hamiltonian, due to the energy correction,
due to the wave function correction,
and due to the hyperfine modification of the 
electron's transition current.
For the high-energy part, we find a sum of two
terms, one of which is due to a second-order 
effect involving the Pauli form factor correction to the 
Coulomb field, and the second of which is an
anomalous magnetic moment correction to the 
hyperfine splitting, evaluated on relativistic
wave functions. The first correction can be split into 
two terms, which involve/do not involve mixing of the 
upper and lower components of the Dirac wave function,
respectively.
Quite surprisingly, the high-energy contribution
can be expressed in closed analytic form,
valid for an arbitrary excited state [see Eq.~\eqref{ThetaH}].
For the Bethe logarithm type corrections
relevant to the low-energy part, a numerical 
approach is indispensable.

Finally, as indicated in Tables~\ref{table1}---\ref{table3},
we also find results for $D$, $F$, and $G$ states
which are of general interest to high-precision 
spectroscopy.

%
% Acknowledgments
%
\section*{Acknowledgments}

This research has been supported by the National
Science Foundation and by a Precision Measurement
Grant from the National Institute of Standards and Technology.

\end{document}